\documentclass[11pt]{article}
\pdfoutput=1

\usepackage{jheppub}
\usepackage{url,comment}
\usepackage{times}
\usepackage{latexsym}
\usepackage[dvipsnames]{xcolor}
\usepackage{graphicx, graphics, amsmath, amssymb, slashed, bbm,bm,amsthm, array}
\usepackage{subfigure}
\usepackage{listings} 
\usepackage{trimclip}
\usepackage{xspace}
\usepackage{rotating}
\usepackage{afterpage}

\usepackage{makecell}
 
\usepackage{multirow}

\usepackage{hyperref}	
\hypersetup{  colorlinks=true,     linkcolor=blue,     filecolor=magenta,          urlcolor=blue}

\newcommand{\nc}{\newcommand}

\nc{\beq}{\begin{equation}}
\nc{\eeq}{\end{equation}}
\nc{\barray}{\begin{eqnarray}}
\nc{\earray}{\end{eqnarray}}
\nc{\barrayn}{\begin{eqnarray*}}
\nc{\earrayn}{\end{eqnarray*}}
\nc{\bcenter}{\begin{center}}
\nc{\ecenter}{\end{center}}
\nc{\mc}{\mathcal}
\nc{\er}[1]{(\ref{eq:#1})}
\nc{\onehalf}{\frac{1}{2}} 
\nc{\partialbar}{\bar{\partial}}
\nc{\psit}{\widetilde{\psi}}
\nc{\Tr}{\mbox{Tr}}
\nc{\hc}{\mbox{H.c.}}
\nc{\ev}{\;\mathrm{eV}}
\nc{\mev}{\;\mathrm{MeV}}
\nc{\gev}{\;\mathrm{GeV}}
\nc{\kev}{\;\mathrm{keV}}
\nc{\tev}{\;\mathrm{TeV}}
\nc{\pev}{\;\mathrm{PeV}}
\nc{\eev}{\;\mathrm{EeV}}

\def\chii0{\chi_i^0}
\def\chij0{\chi_j^0}

\newcommand{\gsim}{\lower.7ex\hbox{$\;\stackrel{\textstyle>}{\sim}\;$}}
\newcommand{\lsim}{\lower.7ex\hbox{$\;\stackrel{\textstyle<}{\sim}\;$}}
\nc{\ttbar}{t\bar t}

\newcommand{\fref}[1]{Fig.~\ref{#1}}

\newcommand{\cref}[1]{Chapter~\ref{#1}}

\graphicspath{{plots/}}

\newcommand{\glossarycolor}{black}

\newcommand{\tower}{\textcolor{\glossarycolor}{tower module}\xspace}
\newcommand{\towers}{\textcolor{\glossarycolor}{tower modules}\xspace}

\newcommand{\Towers}{\textcolor{\glossarycolor}{Tower modules}\xspace}

\newcommand{\SiPM}{\textcolor{\glossarycolor}{SiPM}\xspace} 
\newcommand{\SiPMs}{\textcolor{\glossarycolor}{SiPMs}\xspace} 

\newcommand{\WLSF}{\textcolor{\glossarycolor}{WLSF}\xspace}

\newcommand{\mbar}{\textcolor{\glossarycolor}{bar}\xspace} 

\newcommand{\mbars}{\textcolor{\glossarycolor}{bars}\xspace} 
\newcommand{\mBars}{\textcolor{\glossarycolor}{Bars}\xspace}

\newcommand{\barass}{\textcolor{\glossarycolor}{bar assembly}\xspace} 

\newcommand{\barasss}{\textcolor{\glossarycolor}{bar assemblies}\xspace}

\newcommand{\layer}{\textcolor{\glossarycolor}{tracking layer}\xspace} 
\newcommand{\layers}{\textcolor{\glossarycolor}{tracking layers}\xspace}

\newcommand{\Layers}{\textcolor{\glossarycolor}{Tracking layers}\xspace}

\newcommand{\trackingmodule}{\textcolor{\glossarycolor}{tracking module}\xspace} 
\newcommand{\trackingmodules}{\textcolor{\glossarycolor}{tracking modules}\xspace}

\newcommand{\ceilingtrackingmodule}{\textcolor{\glossarycolor}{ceiling tracking module}\xspace} 
\newcommand{\ceilingtrackingmodules}{\textcolor{\glossarycolor}{ceiling tracking modules}\xspace} 
 
\newcommand{\Ceilingtrackingmodules}{\textcolor{\glossarycolor}{Ceiling tracking modules}\xspace}

\newcommand{\walltrackingmodules}{\textcolor{\glossarycolor}{wall tracking modules}\xspace}

\newcommand{\wallvetolayers}{\textcolor{\glossarycolor}{front wall  veto layers}\xspace}

\newcommand{\wallveto}{\textcolor{\glossarycolor}{front wall veto detector}\xspace}

\newcommand{\columndetector}{\textcolor{\glossarycolor}{column detector}\xspace} 
\newcommand{\columndetectors}{\textcolor{\glossarycolor}{column detectors}\xspace}

\newcommand{\floorvetolayers}{\textcolor{\glossarycolor}{floor veto layers}\xspace}

\newcommand{\floorvetostrips}{\textcolor{\glossarycolor}{floor veto strips}\xspace} 
 
\newcommand{\Floorvetostrips}{\textcolor{\glossarycolor}{Floor veto strips}\xspace}

\newcommand{\floorveto}{\textcolor{\glossarycolor}{floor veto detector}\xspace}

\newcommand{\veto}{\textcolor{\glossarycolor}{veto detector}\xspace}

\newcommand{\barlength}{\textcolor{\glossarycolor}{2.35~m}\xspace}
\newcommand{\barwidth}{\textcolor{\glossarycolor}{3.5~cm}\xspace}
\newcommand{\barthickness}{\textcolor{\glossarycolor}{1~cm}\xspace}

\title{
MATHUSLA: An External Long-Lived Particle Detector to Maximize the Discovery Potential of the HL-LHC
\\ \vspace*{3mm}
\textnormal{\normalsize{\textit{Input to the 2026 update of the European Strategy for Particle Physics, \today}}}
}

\author{
\includegraphics[width=4cm]{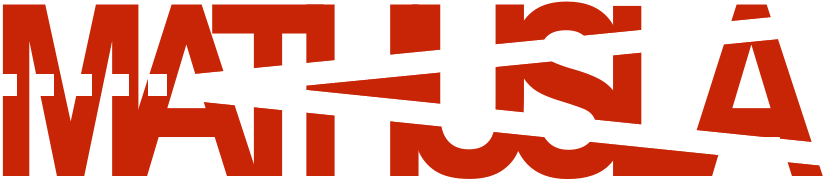}
\\
{\normalfont 
\href{https:mathusla-experiment.web.cern.ch}{\texttt{mathusla-experiment.web.cern.ch}}
\\
\vspace{5mm}}}

\author[1]{\hspace*{-1mm}Branden   Aitken,}
\author[2]{Cristiano  Alpigiani,}
\author[3]{Juan Carlos   Arteaga-Vel\'azquez,}
\author[4]{Mitchel   Baker,}
\author[5]{Kincso   Balazs,}
\author[6]{Jared   Barron,}
\author[7]{Brian   Batell,}
\author[8]{Austin   Batz,}
\author[9]{Yan   Benhammou,}
\author[5]{Tamara Alice  Bud,}
\author[10]{Karen Salom\'e   Caballero-Mora,}
\author[11]{John Paul   Chou,}
\author[12]{David   Curtin,}
\author[5]{Albert   de Roeck,}
\author[12]{Miriam   Diamond,}
\author[13]{Mariia   Didenko,}
\author[14, 15]{Keith R.    Dienes,}
\author[16]{William  Dougherty,}
\author[5]{Liam Andrew   Dougherty,}
\author[17]{Marco   Drewes,}
\author[11]{Sameer   Erramilli,}
\author[9]{Erez   Etzion,}
\author[18]{Arturo   Fern\'andez T\'ellez,}
\author[11]{Grace  Finlayson,}
\author[19]{Oliver   Fischer,}
\author[20]{Jim   Freeman,}
\author[5]{Jonathan   Gall,}
\author[2]{Ali   Garabaglu,}
\author[21]{Bhawna   Gomber,}
\author[11]{Stephen Elliott   Greenberg,}
\author[12, 22]{Jaipratap Singh  Grewal,}
\author[1]{Zoe   Hallman,}
\author[11]{Bahgat   Hassan,}
\author[23]{Yuekun   Heng,}
\author[12]{Keegan   Humphrey,}
\author[1]{Trystan   Humphrey,}
\author[8]{Graham  D.  Kribs,}
\author[12]{Alex   Lau,}
\author[12]{Jiahao   Liao,}
\author[15]{Zhen   Liu,}
\author[24]{Giovanni   Marsella,}
\author[5]{Matthew   McCullough,}
\author[25]{David   McKeen,}
\author[6]{Patrick   Meade,}
\author[1]{Caleb  Miller,}
\author[9]{Gilad   Mizrachi,}
\author[10]{O. G.   Morales-Olivares,}
\author[25]{David   Morrissey,}
\author[26]{Abdulrahman Ahmed   Morsy,}
\author[5]{John   Osborn,}
\author[12]{Gabriel  Owh,}
\author[27]{Michalis  Panagiotou,}
\author[2]{Mason   Proffitt,}
\author[12]{Runze  Ren,}
\author[4]{Steven H.   Robertson,}
\author[18]{Mario   Rodr\'iguez-Cahuantzi,}
\author[1]{Heather   Russell,}
\author[13]{Victoria   S\'anchez,}
\author[27]{Halil  Saka,}
\author[1]{Mamoksh   Samra,}
\author[28]{Rodney   Schnarr,}
\author[29]{Jessie   Shelton,}
\author[9]{Yiftah   Silver,}
\author[30]{Daniel   Stolarski,}
\author[31]{Martin A.   Subieta Vasquez,}
\author[32]{Sanjay Kumar   Swain,}
\author[11]{Steffie Ann   Thayil,}
\author[33]{Brooks   Thomas,}
\author[13]{Emma   Torro,}
\author[34]{Yuhsin   Tsai,}
\author[1]{Bennett   Winnicky-Lewis,}
\author[9]{Igor  Zolkin,}
\author[13]{Jose   Zurita}

\affiliation[1]{University of Victoria, Canada}
\affiliation[2]{University of Washington, Seattle, USA}
\affiliation[3]{Universidad Michoacana de San Nicol\'as de Hidalgo, Mexico (UMSNH)}
\affiliation[4]{University of Alberta, Canada}
\affiliation[5]{CERN, Switzerland}
\affiliation[6]{YITP Stony Brook, USA}
\affiliation[7]{University of Pittsburgh, USA}
\affiliation[8]{University of Oregon, USA}
\affiliation[9]{Tel Aviv University, Israel}
\affiliation[10]{Universidad Aut\'onoma de Chiapas, Mexico (UNACH)}
\affiliation[11]{Rutgers, the State University of New Jersey, USA}
\affiliation[12]{University of Toronto, Canada}
\affiliation[13]{Instituto de F\'isica Corpuscular (CSIC-UV), Valencia, Spain}
\affiliation[14]{University of Arizona, USA}
\affiliation[15]{University of Maryland, USA}
\affiliation[16]{Kenmore, Washington, USA}
\affiliation[17]{Universit\'{e} catholique de Louvain, France}
\affiliation[18]{Benem\'erita Universidad Aut\'onoma de Puebla, Mexico (BUAP)}
\affiliation[19]{Liverpool U., UK}
\affiliation[20]{Fermi National Accelerator Laboratory (FNAL), USA}
\affiliation[21]{Hyderabad University, India}
\affiliation[22]{University of California San Diego, USA}
\affiliation[23]{Institute of High Energy Physics, Beijing}
\affiliation[24]{Universit\`a degli di Studi di Palermo, Palermo, Italy}
\affiliation[25]{TRIUMF, Canada}
\affiliation[26]{Ain Shams University, Cairo, Egypt}
\affiliation[27]{University of Cyprus, Cyprus}
\affiliation[28]{Carleton University, Canada}
\affiliation[29]{University of Illinois Urbana-Champaign, USA}
\affiliation[30]{Carleton Unversity, Canada}
\affiliation[31]{Instituto de Investigaciones F\'isicas (IIF), Observatorio de F\'isica C\'osmica de \^a Chacaltaya\^a, Universidad Mayor de San Andr\'es (UMSA)}
\affiliation[32]{National Institute of Science Education and Research, HBNI, Bhubaneswar, India}
\affiliation[33]{Lafayette College, USA}
\affiliation[34]{University of Notre Dame, USA}

\emailAdd{mathusla.experiment@cern.ch}

\abstract{
We present the current status of the 
MATHUSLA (MAssive Timing Hodoscope for Ultra-Stable neutraL pArticles) long-lived particle (LLP) detector at the HL-LHC,
covering the design, fabrication and installation at CERN Point 5.
MATHUSLA40 is a 40~m-scale  detector with an air-filled decay volume that is instrumented with scintillator tracking detectors, to be located near CMS.
Its large size, close proximity to the CMS interaction point and about 100~m of rock shielding from LHC backgrounds allows it to detect LLP production rates and lifetimes that are  one to two orders of magnitude beyond the  ultimate reach of the LHC main detectors. 
This provides unique sensitivity to  many LLP signals that are highly theoretically motivated, due to their connection to the hierarchy problem, the nature of dark matter, and baryogenesis. 
Data taking is projected to commence with the start of HL-LHC operations.
We summarize the new 40m design for the detector that was recently presented in the MATHUSLA Conceptual Design Report, alongside new realistic background and signal simulations that demonstrate high efficiency for the main target LLP signals in a background-free HL-LHC search. 
We argue that MATHUSLA's uniquely robust expansion of the HL-LHC physics reach is a crucial ingredient in CERN's mission to search for new physics and characterize the Higgs boson with precision.
}

\notoc
\begin{document}

\maketitle
\thispagestyle{empty}
\newpage

\section{Introduction and Executive Summary}

\pagestyle{plain}

As the High Luminosity (HL) era of the Large Hadron Collider (LHC) approaches, it becomes ever more pertinent to ensure that the physics return on the world's massive investment in the decades-long LHC program is maximized. 
\textbf{Neutral Long-Lived particles (LLPs) at the weak scale are a highly theoretically motivated possibility for physics beyond the Standard Model (BSM)}, both from bottom-up considerations and as top-down predictions of many proposed theories of Dark Matter, Baryogenesis, and solutions to the Hierarchy Problem (see Ref.~\cite{Curtin:2018mvb} for a comprehensive review). 
As a result, the LLP search program at the LHC has undergone dramatic development in recent years~\cite{Alimena:2019zri}, but even with upcoming upgrades, \textbf{the HL-LHC main detectors suffer from  and complex backgrounds and trigger limitations~\cite{Alimena:2021mdu}}.

\begin{figure}
    \centering
   \includegraphics[width=0.7 \textwidth]{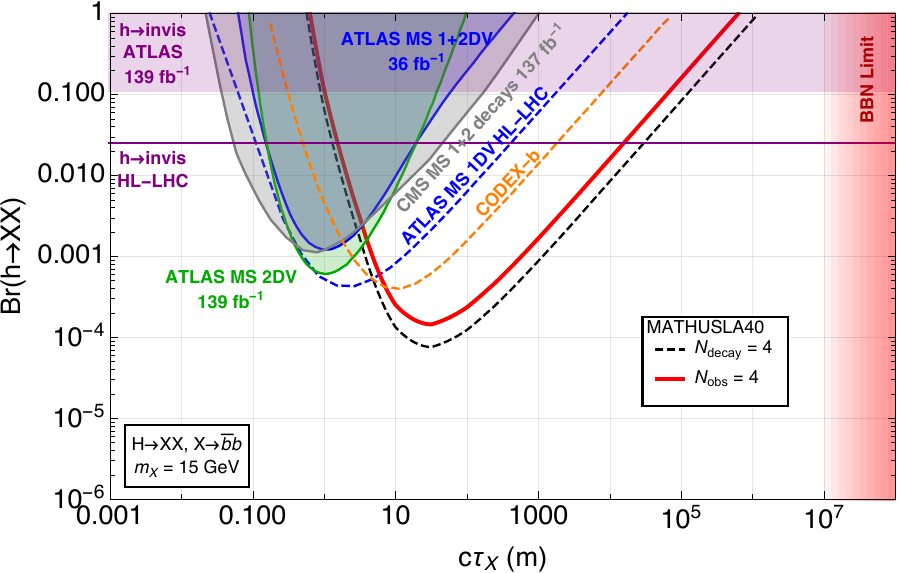}
    \caption{
    Sensitivity of the 40m MATHUSLA40 detector to hadronically decaying LLPs produced in exotic Higgs boson decays. The solid red curve shows the exclusion reach from a realistic background-free search for DVs with with 49\% signal reconstruction efficiency after background vetoes are applied. For comparison, the dashed black curve shows idealized reach for 4 decays in the decay volume.
    We also show the current $\mathrm{Br}(h \to \mathrm{invis})$ limit from ATLAS~\cite{ATLAS:2023tkt} (purple shading) and  the HL-LHC projection~\cite{Dainese:2019rgk} (purple line); current ATLAS constraints from searches for 1 or 2 DVs (blue shading) and 2 DVs (green shading) in the muon system~\cite{ATLAS:2018tup, ATLAS:2022gbw};
    current CMS constraints from searches for 1 or 2 LLP decays in the muon system~\cite{CMS:2024bvl} (gray shading);
    projections for an ATLAS 1DV search in the muon system at the HL-LHC~\cite{Coccaro:2016lnz} (blue dashed);
    and the idealized sensitivity of CODEX-b~\cite{Aielli:2019ivi} (orange dashed).    }
    \label{f.higgs}
\end{figure}

\textbf{One of the most important such blind-spots are LLPs in the 10-100 GeV mass range that decay hadronically, which arise in a wide variety of scenarios.}
Discovering these LLPs is the primary physics case of the 
The MATHUSLA (MAssive Timing Hodoscope for Ultra-Stable neutraL pArticles) proposal for a large external LLP detector on the surface next to CMS.
\textbf{MATHUSLA40 would extend the LLP reach of the main detectors by orders of magnitude~\cite{Chou:2016lxi, MATHUSLA:2018bqv, MATHUSLA:2020uve, MATHUSLA:2022sze, Aitken:2025cjq}, see Fig.~\ref{f.higgs}, and is the most robust proposal to detect these elusive signals and  maximize the discovery potential of the HL-LHC.}

\begin{figure}
\begin{center}
\includegraphics[width=0.7\textwidth]{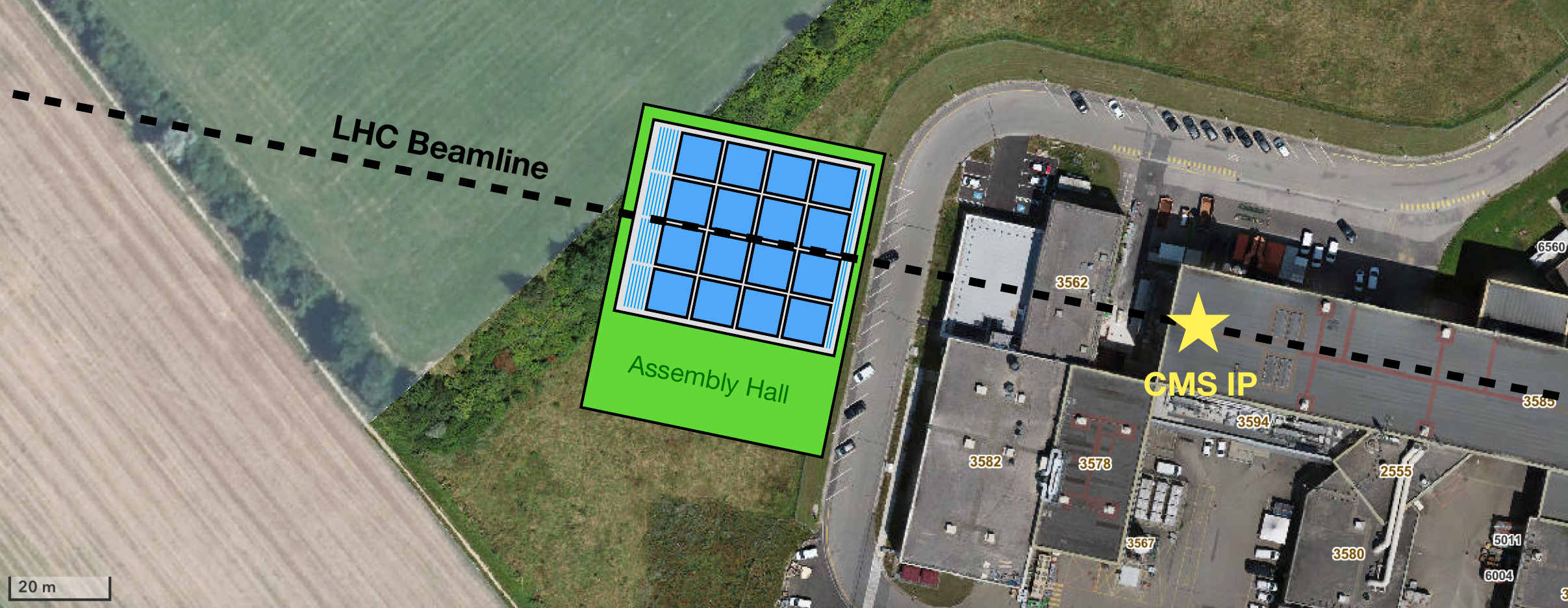}
\end{center}
\caption{Location of proposed MATHUSLA40 detector at the CMS site. 
}
\label{fig:layout_P5}
\end{figure}

\begin{figure}
\begin{center}
\includegraphics[width=0.9\textwidth]{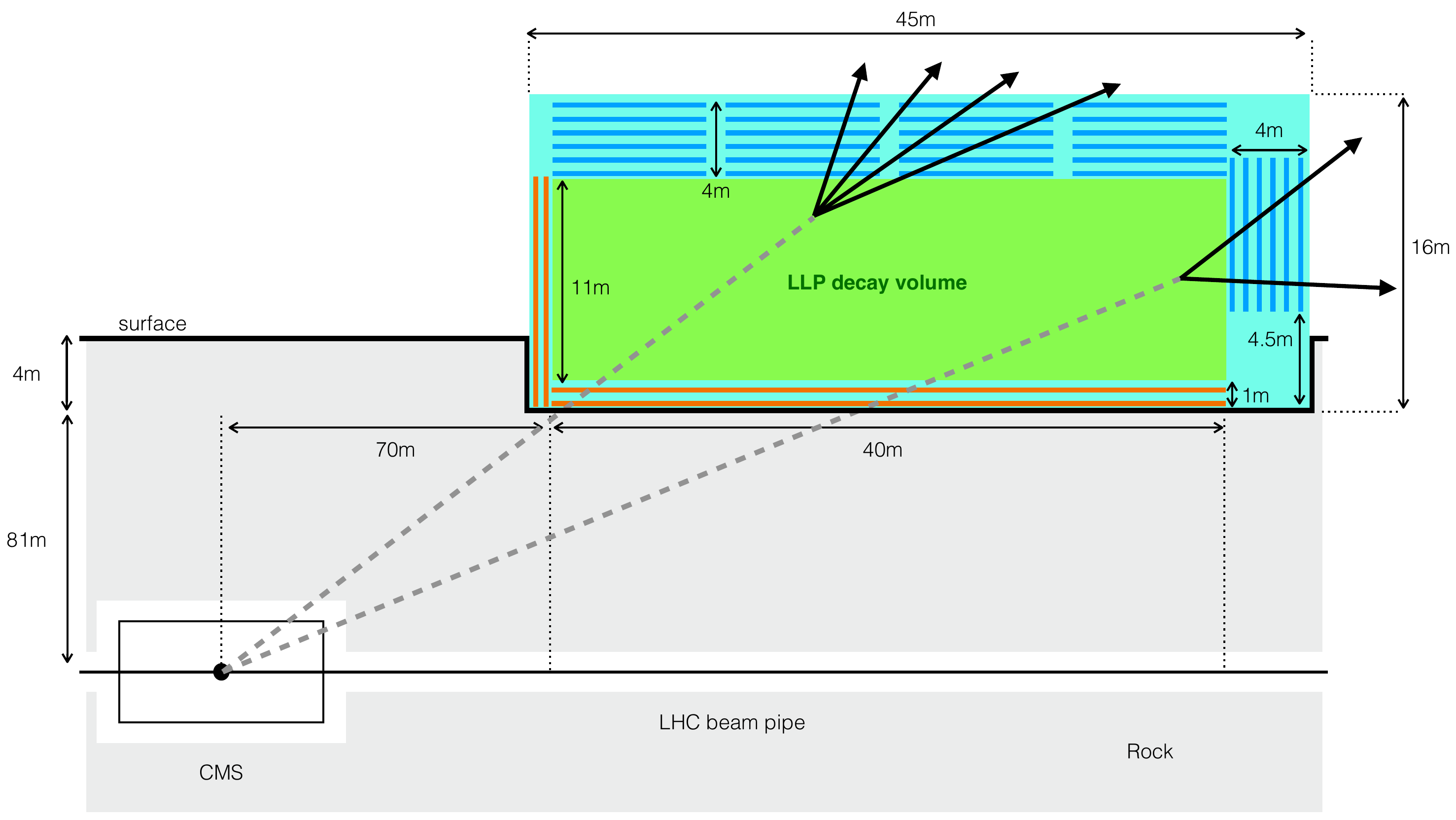}
\end{center}
\caption{
Schematic MATHUSLA40 geometry relative to the CMS collision point. 
LLPs (gray dashed) can decay into SM charged states (black arrows) in the 
$\sim 40~\mathrm{m} \times 40~\mathrm{m} \times 11~\mathrm{m}$ \textbf{LLP decay volume (green)}
and be reconstructed as displaced vertices by the \textbf{\trackingmodules (blue)}.
Both wall and ceiling \trackingmodules consist of six $(9~\mathrm{m})^2$ \layers separated by 80cm for a total thickness of 4m, arranged in a $4 \times 4$ grid on the ceiling and a row of 4 on the rear wall, with neighboring \trackingmodules separated by $\sim 1~\mathrm{m}$. 
Rejection of LHC muon and cosmic ray backgrounds is aided by the  double-layer \textbf{\veto} (orange), comprising the
\textbf{\wallveto} and \textbf{\floorveto}, which is close to hermetic for LHC muons and downward traveling cosmics. (The realistic structure of the \floorveto has been simplified for this illustration.) For consistency, we show the modest amount of surface excavation (at most 4m below grade) that may be required to fit the MATHUSLA40 detector into an experimental hall that conforms to local building height restrictions.}
\label{fig:mathuslacms}
\end{figure}

MATHUSLA is dedicated to finding exotic long-lived particles (LLPs) produced in $pp$ collisions in the CMS detector at Point 5.
Its approximately 45 m $\times$ 50 m footprint\footnote{Compared to previous MATHUSLA publications, this is a smaller detector geometry to bring the scale of the project more in line with realistic future funding envelopes at CERN, Europe and North America.} will be situated on CERN-owned, available land adjacent to the CMS complex (\fref{fig:layout_P5}). The basic detector design is simple in principle, consisting of an empty LLP decay volume  that is instrumented with \layers (\fref{fig:mathuslacms}). LLP decays into charged, Standard Model (SM) particles can be reconstructed as displaced vertices (DVs) and distinguished from backgrounds using both their direction of travel and a variety of other timing and geometric criteria. 
Detailed simulations~\cite{Aitken:2025cjq} confirm MATHUSLA40's ability to conduct \textbf{background-free searches for its primary LLP physics targets} with high signal efficiency.
However, even more crucial to MATHUSLA's success is the fact that the extremely dominant cosmic ray backgrounds can be carefully studied \emph{in situ} during the 50\% beam-off duty cycle of the HL-LHC, ensuring that the complete vetoing of backgrounds can be verified without any contamination by potential LLP signals. \textbf{In the event of a positive LLP detection, MATHUSLA40 can therefore robustly claim discovery of a new fundamental particle.} 
Furthermore, MATHUSLA40 can supply a trigger signal to CMS to ensure that the $pp$ collision that produced the LLP is recorded. \textbf{Subsequent correlated off-line analyses can then diagnose many features of the underlying LLP model} with as little as 10 observed decays~\cite{Barron:2020kfo}.

The collaboration recently completed the MATHUSLA40 Conceptual Design Report~\cite{Aitken:2025cjq}, representing significant progress towards the realization of this proposal.
\textbf{The timeline of the HL-LHC makes European Support for the proposal crucial in the coming years, to ensure MATHUSLA40 is constructed in time to take data at the HL-LHC and maximize the unique discovery potential of the world's highest energy proton collisions at high intensity.}

\section{MATHUSLA40 Detector Proposal}

\begin{figure}[]
\begin{center}
\includegraphics[width=0.8\textwidth]{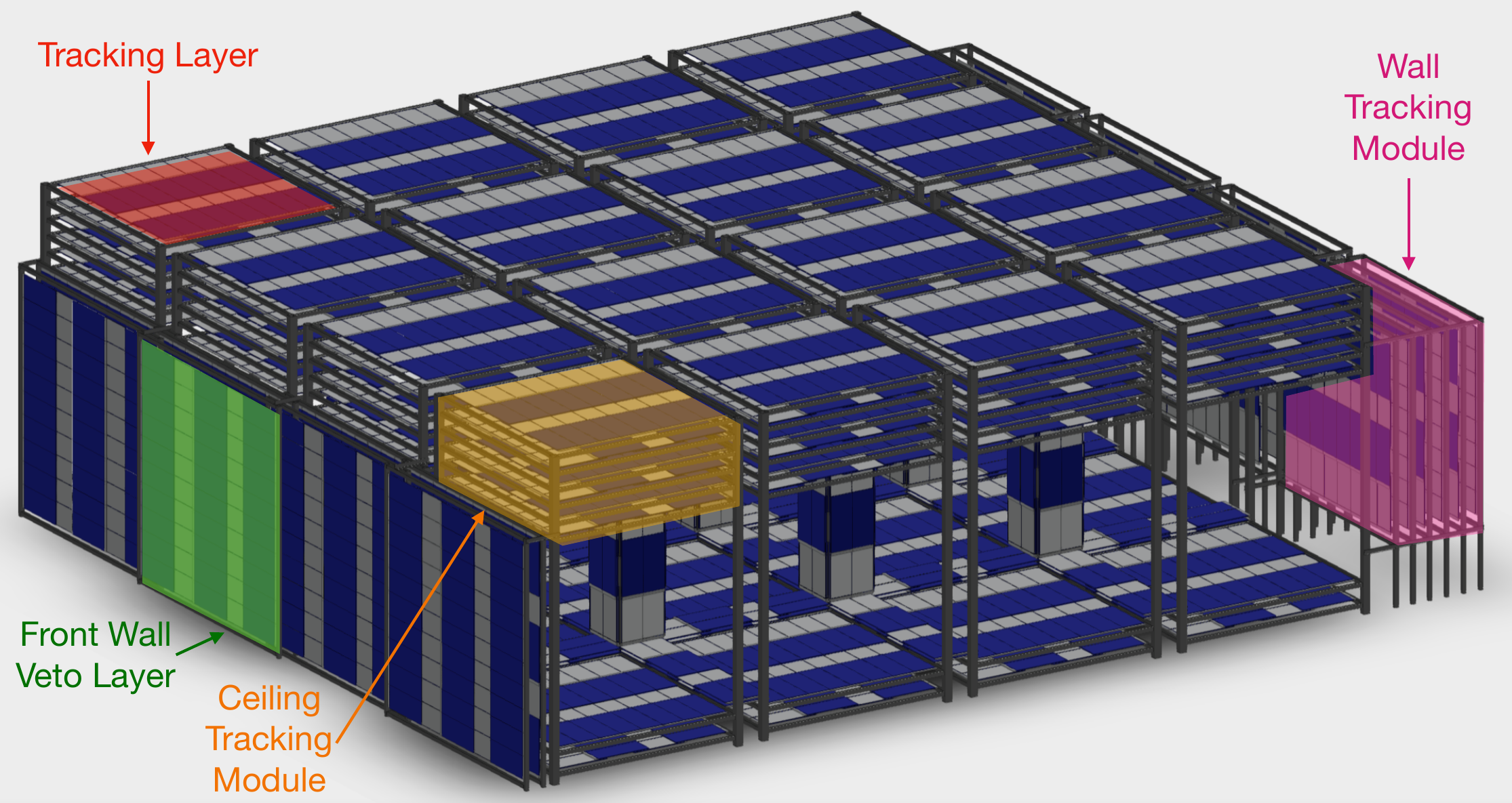}
\\ \vspace{3mm}
\includegraphics[width=0.8\textwidth]{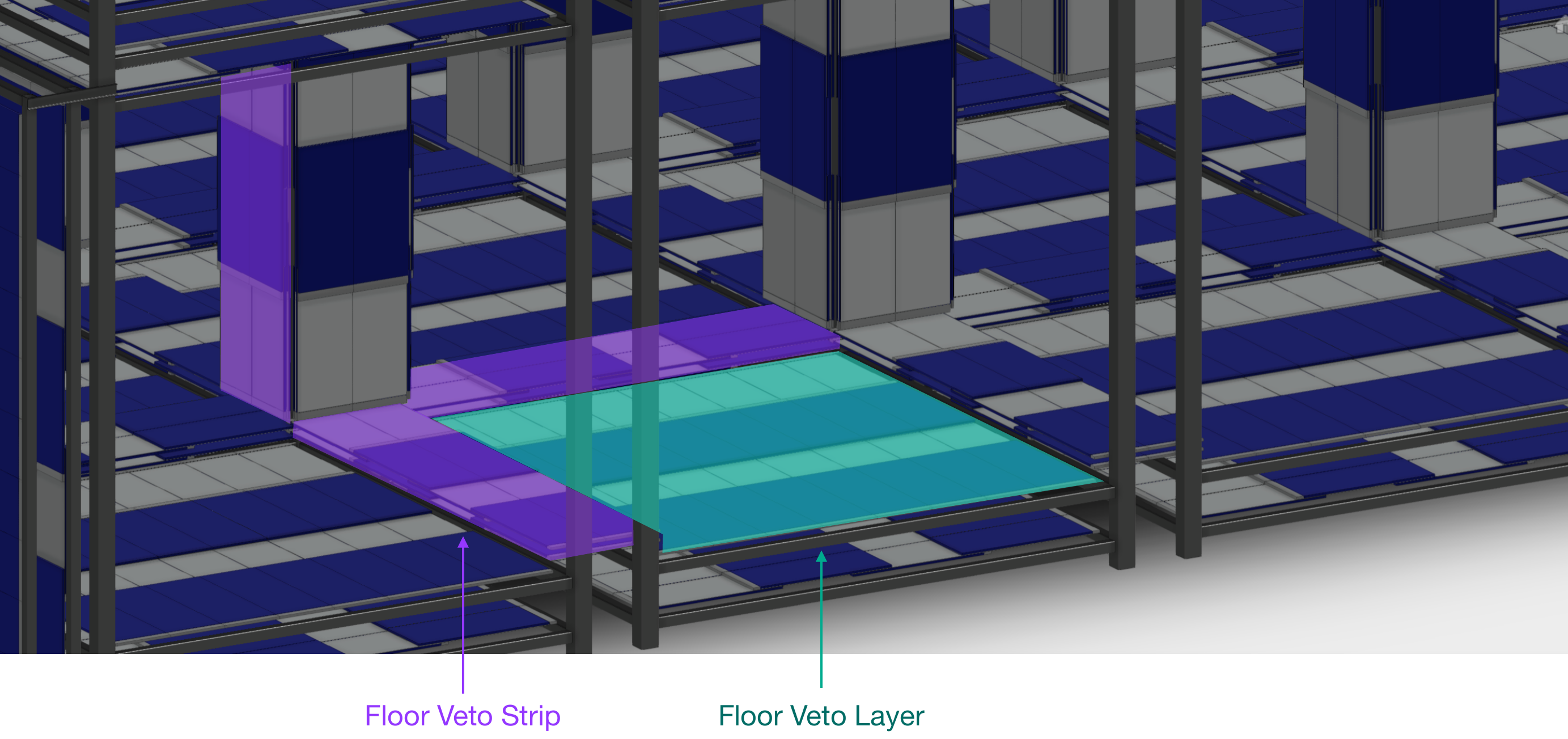}
\end{center}
\vspace*{-5mm}
\caption{
\emph{Top:}
Engineering concept for a realistic MATHUSLA40 detector structure. 
16 \Ceilingtrackingmodules, and 4 \walltrackingmodules in the rear, are each comprised of 6 \layers with 80cm separation. The \ceilingtrackingmodules are mounted in \towers which are arranged in a $4 \times 4$ grid, with 1m separation for maintenance access. 
The \wallveto is comprised of $(11.2~\mathrm{m})^2$ \wallvetolayers, 8 of which are arranged in 2 rows of 4 with overlap to provide hermetic coverage. 
\emph{Bottom:}
The \floorveto comprises \floorvetolayers, identical to those in the \ceilingtrackingmodules and mounted 0.5m and 1m above the floor in the \towers, 
as well as $\sim$ 2.3m $\times$ 9m \floorvetostrips, which cover the gaps between \towers. Four vertical \floorvetostrips also each constitute a single \columndetector, instrumenting the vertical support columns. 
In combination, this \veto system is hermetic for LHC muons and downward cosmics.
}
\label{fig:MATHUSLAstructure}
\end{figure}

\begin{table}[h!]
    \centering
    \caption{Summary of several attributes of the MATHUSLA40 detector benchmark design~\cite{Aitken:2025cjq}. 
    }
    \hspace*{-3mm}
    \begin{tabular}{|m{0.3 \textwidth}|m{0.68\textwidth}|}
    \hline
    Distance from CMS IP
    &
    82-93m vertical, 70-110m horizontal along beam axis
    \\ \hline
    Detector volume & $\sim$ $40~\mathrm{m} \times 45~\mathrm{m} \times 16~\mathrm{m} $ 
    \\ \hline
    Decay volume
    &
    $\sim$ $40~\mathrm{m} \times 40~\mathrm{m} \times 11~\mathrm{m}$ 
    \\ \hline
    Number of tracking modules
    & 
    20 total: a grid of $4 \times 4$ \towers each has a \ceilingtrackingmodule, and 4 \walltrackingmodules are mounted on the rear wall.
    \\ \hline
    Tracking module Dimensions
    &
    9~m $\times$ 9~m, height $\sim$ 4~m 
    \\ \hline
    \Layers
    & 
    6 in ceiling (top 4m, 0.8m apart) and 6 in rear wall (starting $\sim$ 4.5m above the floor, also 0.8m apart).\\ \hline
    Hermetic wall detector
    & 
    Double layer in wall facing IP to detect LHC muons.
 \\ \hline
    Hermetic floor detector
    & 
    2 \floorvetolayers at heights 0.5m and 1m in each of the 16 \towers, 24 $(9~\mathrm{m} \times 2.8~\mathrm{m})$ \floorvetostrips to cover gaps between \towers, and 9 \columndetectors each utilizing 4 vertical \floorvetostrips to cover the vertical support columns.
     \\ \hline
    Detector technology
    & Extruded plastic scintillator \mbars, \barwidth wide, \barthickness thick, \barlength  long, arranged in alternating orientations with each vertical \layer. \mBars are threaded with wavelength-shifting fibers connected to \SiPMs. 
    \\ \hline
    Number of \barasss
    &
    $6224$, 32 channels each
    \\ \hline
    Number of Channels
    &
    $\sim 2\times 10^5$ \SiPMs
    
    \\ \hline
    Tracking resolution
    & 
    $\sim$~1~ns timing resolution;
    $\sim$ 1~cm (15~cm) along transverse (longitudinal) direction of scintillator bar.
     \\ \hline
    Trigger
    & 
    $3 \times 3$ groups of \trackingmodules perform simplified tracking/vertexing to trigger on upwards-traveling tracks and vertices. 
    Corresponding time stamps flag regions of MATHUSLA datastream for full reconstruction and permanent storage. 
    MATHUSLA can also send hardware trigger signal to CMS to record LLP production event.
     \\ \hline
    Data rate
    & 
    Each \trackingmodule and section of \floorveto detector associated with each \tower produces $\lesssim$~0.6 TB/day. (The \wallveto data rate is a small addition.)  Less than 0.1\% of full detector data will be selected for permanent storage using a trigger system, corresponding to about 8~TB/year. 
    \\ \hline
    \end{tabular}
    \label{t.summary}
\end{table}

An overview of the engineering concept for a realistic implementation of the proposed MATHUSLA40 detector is shown in Fig.~\ref{fig:MATHUSLAstructure}.
Key attributes of the design are summarized in Table~\ref{t.summary}.

In order to improve on the LLP sensitivity of the LHC main detectors by orders of magnitude, while also complying with maximum building height regulations near LHC Point 5 and minimizing potentially expensive excavation, MATHUSLA40's  \textbf{LLP decay volume} has a footprint of $\sim (40~\mathrm{m})^2$ and height of $\sim$ 11~m, with about a meter on the bottom being taken up by the \textbf{\floorveto} and 4~m on top occupied by the \textbf{\ceilingtrackingmodules}. Additional \textbf{\walltrackingmodules} on the back wall relative to the LHC IP greatly enhance signal reconstruction efficiency for LLPs decaying in the rear of the detector, while a \textbf{\wallveto} enhances rejection of LHC muon backgrounds. 
The modular structure of the  detector makes a staged installation of the tracking modules possible. 

Each \trackingmodule, to be installed in the ceiling or rear wall, is comprised of 6 \layers with an area of $\sim (9~\mathrm{m})^2$, separated by 80~cm for a total height of 4~m. 
Each \ceilingtrackingmodule is mounted in a \tower with four vertical supports, which also support the components of the \floorveto. 
\Towers are arranged in a $4 \times 4$ grid to cover the entire decay volume, and are separated by 1~m-wide gaps.
The gaps have little impact on signal efficiency but are crucial for allowing maintenance access. 
The rear \walltrackingmodules are similarly arranged in a row with 1~m gaps between them. 

\begin{figure}[]
\begin{center}
\includegraphics[height=3.3cm]{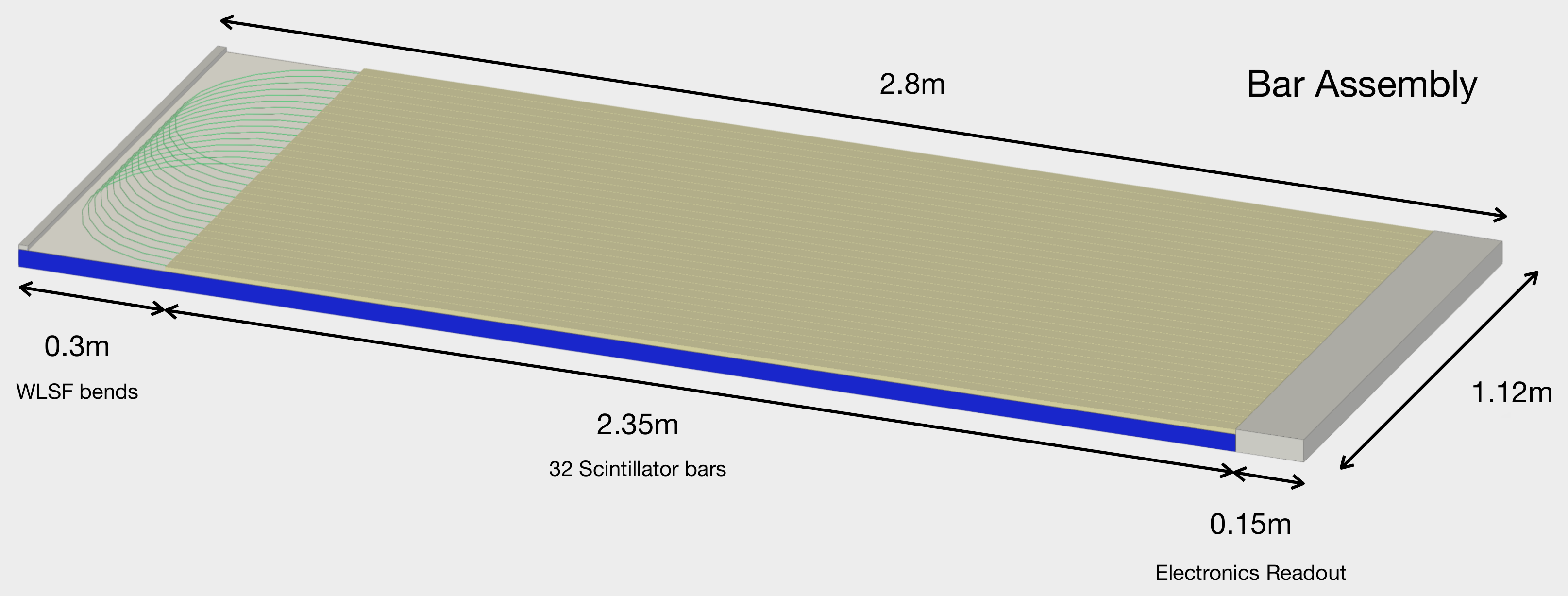}
\includegraphics[height=3.3cm]{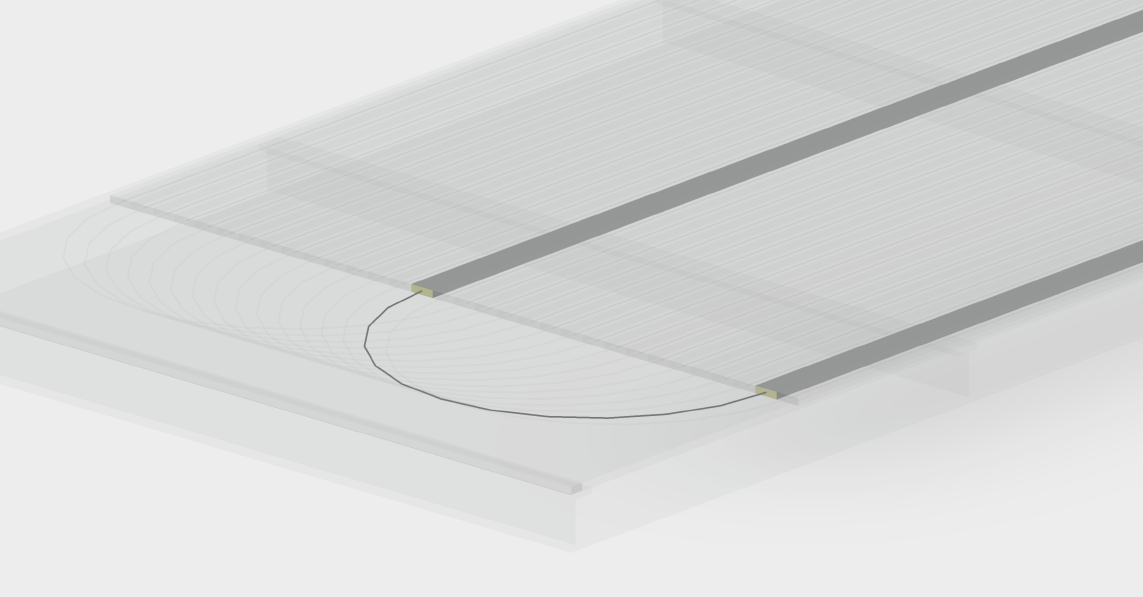}
\end{center}
\caption{Details of the \barass made of 32 scintillator \mbars. 
Left: overview of the \barass. The electronics readout box contains the Silicon Photomultipliers (\SiPMs) and electronics board. At the other end, the \WLSF "bends" as the fibers go down one bar and back through another bar. In this way all \SiPMs for the \barass are on the same end. Right: details of the \WLSF bend region.}
\label{fig:FNAL_extrusions}
\end{figure}

The 6 \layers in each tracking module are comprised of \textbf{scintillating \mbars} of \barwidth width that are arranged with alternating transverse orientation to their neighboring \layers. 
This configuration provides position and timing coordinates of charged particles resulting from the decay of LLPs in the \mbox{MATHUSLA40} detector decay volume with $\sim 1~\mathrm{ns}$ timing and $\sim$~cm transverse spatial resolution. 
The \layers in each tracking module on the ceiling or wall are also used as {\bf trigger layers} in addition to providing tracking information.
%
The number of layers was also optimized with full simulations to ensure the primary physics target  can be searched for with effectively zero background.

Each of the two layers of the \wallveto  is implemented by a slightly staggered arrangement of four $(11.2~\mathrm{m})^2$ {\wallvetolayers} to provide hermetic coverage for the full 40m width of the front wall. 
The \floorveto is comprised of \floorvetolayers and \floorvetostrips. 
The \floorvetolayers are identical to \layers in the \trackingmodules, and are mounted at heights of 0.5m and 1m above the floor in the tower modules to cover the majority of the floor area.
\Floorvetostrips have physical dimensions of 9m $\times$ 2.8m and each provide about 9m $\times$ 2.3m of double-layer sensor coverage. They are are mounted horizontally above the \floorvetolayers to cover the gaps between \towers, and are also mounted vertically around the support columns to constitute \columndetectors that enclose the space at the corners of the \towers. 
In addition to making the floor detector hermetic with respect to cosmics, these \columndetectors also provide explicit material veto capabilities for inelastic cosmic ray interactions in the support column.

The basic sensor building block of MATHUSLA is the \textbf{\barass}, shown in Figure~\ref{fig:FNAL_extrusions}. A single \barass comprises 32 scintillator \mbars of length \barlength, width \barwidth and thickness \barthickness, providing 2.35~m $\times$ 1.12~m area of sensor coverage with an approximate total physical size of 2.8~m $\times$ 1.12~m. 
Each \mbar is extruded with a hole at the center into which a $1.5$~mm diameter wavelength-shifting fiber (\WLSF) is inserted and connected at each end to a Silicon Photomultiplier (\SiPM).  The coordinate along the length of the bar is determined by the differential time measurement of the two ends of the bar, which has a resolution of $\sigma \sim \pm$ $15$~cm. The width of the bar determines the corresponding transverse coordinate with $\sigma \sim \pm 1$ cm. A $\sim 5$m long \WLSF is threaded through two nearby bars, with a $180^\circ$ bend at one end, so that \SiPM signals can be recorded only at one end of the \mbars
Each \barass requires 32  \SiPMs, one for each bar. The electronics readout box at one end of the bar assembly can be mounted flush with either the top or bottom of the bar assembly (from the perspective of Figure~\ref{fig:FNAL_extrusions}), allowing \barasss to be joined with minimal vertical gap in various configurations. 
The bars and electronics are attached to an aluminum strongback plate, resulting in a total thickness of $\sim$ 5cm and making each \barass self-supporting if mounted at the edges.
Bar assemblies are then joined in various ways to make up all the different types of sensor planes used in MATHUSLA.

\begin{figure}[]
\begin{center}
\includegraphics[width=\textwidth]{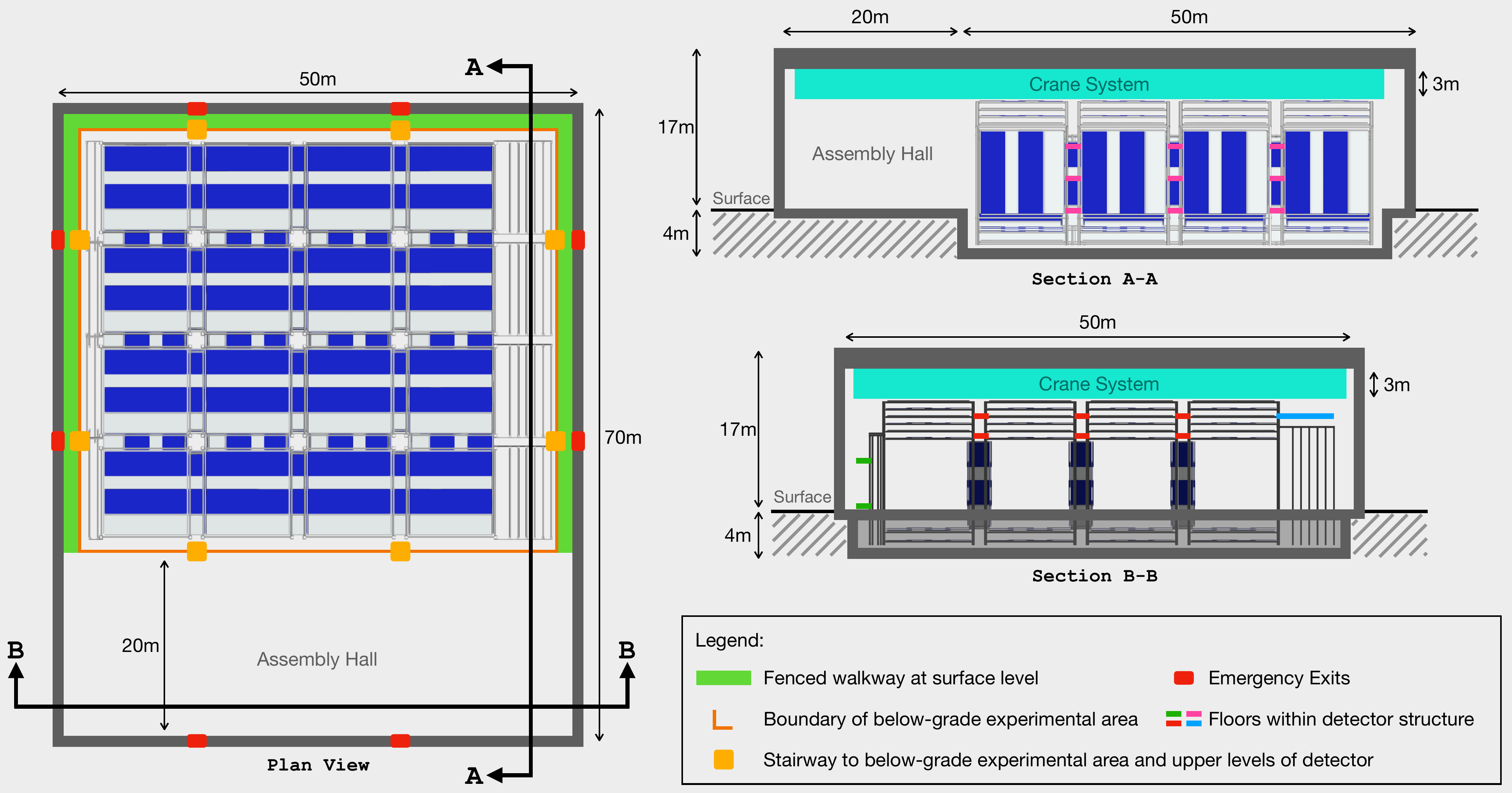}
\end{center}
\caption{
Sketch of the civil engineering concept for the MATHUSLA40 experimental hall (all indicated dimensions approximate). The total footprint of the building is 50m $\times$ 70m, with a total exterior height of 17m. The detector (footprint 40m $\times$ 45m, total height 16m) is situated in a below-grade experimental area (within orange boundary in top view), requiring at most 4m of excavation. This ensures sufficient vertical space for an interior crane system utilizing low-headrooom hoists (cyan) below the roof. Detector components are prepared for installation in the adjacent 20m wide assembly hall, then lifted into place with the cranes. 
} 
\label{fig:civil_engineering_concept}
\end{figure}

The proposed location of MATHUSLA40 near CMS is shown in Fig.~ \ref{fig:layout_P5}. An enclosing experimental hall will need to be constructed, see Fig.~\ref{fig:civil_engineering_concept}, which includes an assembly area adjacent to the detector, in addition to the $\sim$ 45m $\times$ 50m total physical footprint of the detector itself.
The structure would be located on the surface near the CMS Interaction point (IP), fitting entirely on CERN-owned land and allowing MATHUSLA40 to be centered on the LHC beamline. The site allows for the detector to be as close as 68~m horizontally from the IP, which is marked with a yellow star in Figure \ref{fig:layout_P5}, underlying our benchmark assumption that the actual decay volume has a horizontal distance of 70~m from the IP.

\section{Scientific Objectives}

MATHUSLA40 will add a crucial capability to the LHC physics program, beyond the current capabilities of the LHC main detectors.  As discussed in the MATHUSLA physics case white paper~\cite{Curtin:2018mvb},  LLP signals are broadly motivated and ubiquitous in BSM scenarios. Their discovery and subsequent characterization could resolve many fundamental mysteries of high energy physics, including the Hierarchy Problem, the nature of Dark Matter, the origin of the Universe's matter-antimatter asymmetry,  neutrino masses and the strong CP problem.

The physics goal of MATHUSLA is the search for electromagnetically neutral LLPs produced at the HL-LHC. The primary physics target for MATHUSLA 
is informed by the blind-spots of the HL-LHC main detectors~\cite{Chou:2016lxi, Alimena:2019zri, Alimena:2021mdu, ATLAS:2012av, ATLAS:2018tup,ATLAS:2019jcm,ATLAS:2022gbw}, which 
severely limit the sensitivity of many LLP searches, especially for \emph{medium-mass} LLPs (10 to few 100 GeV) that decay hadronically, or \emph{low-mass} LLPs ($\lesssim$ few GeV) of any decay mode. 
While the latter will be well-covered by the recently approved SHiP experiment~\cite{Beacham:2019nyx}, the former blind spot 
requires the full collision energy and luminosity of the HL-LHC as well as a background-free environment to resolve. Maximizing the discovery potential of the HL-LHC in this fashion is MATHUSLA's mission.

\textbf{The main physics target of MATHUSLA is therefore hadronically decaying LLPs in the 10 to few 100 GeV mass range. This is a highly motivated and very general new physics scenario that specifically occurs in many theories, including the neutral naturalness solutions to the hierarchy problem~\cite{Chacko:2005pe, Craig:2015pha, Curtin:2015fna, Batz:2023zef}) and exotic Higgs boson decays to LLPs in any hidden sector in general~\cite{Curtin:2013fra}. This also makes MATHUSLA a crucial component of the precision Higgs physics program at the HL-LHC.} 
The absence of backgrounds and trigger limitations allows MATHUSLA to probe LLP production rates \emph{1-2 orders of magnitude} smaller than the main detector at long lifetimes.

The sensitivity of MATHUSLA40 is shown in \fref{f.higgs} for $m_{LLP} = 15~\mathrm{GeV}$. Higher masses are similar, with a shift of the sensitivity curve in lifetime corresponding to the average LLP boost.
The dashed line shows the contour of 4 decays in the MATHUSLA40 volume $(N_\mathrm{decay} = 4)$. Applying the realistic reconstruction efficiency for a background-free LLP search yields the solid red sensitivity curve $(N_\mathrm{obs} = 4)$. 
It is evident that the $(40~\mathrm{m})^2$  MATHUSLA40 design can probe LLP lifetimes and production rates 1-2 orders of magnitude beyond the reach of the HL-LHC main detector searches or the proposed CODEX-b detector.

\begin{figure}
\centering
\includegraphics[width=0.7 \textwidth]{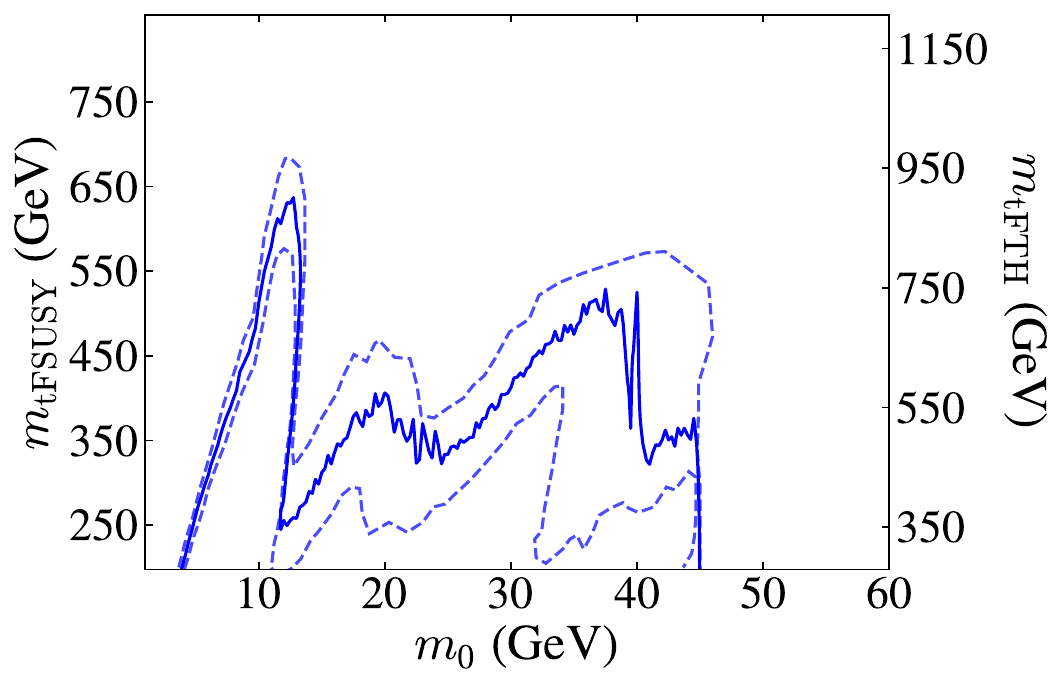}
\caption{
Reach of the 40m MATHUSLA design in a simplified parameter space of Neutral Naturalness, generated using the dark glueball Monte Carlo from~\cite{Batz:2023zef}. Dark glueballs, the lightest of which has mass $m_0$, are produced in exotic Higgs decays which undergo dark Lund-String hadronization. The effective higgs coupling to dark gluons, which also allows glueballs to decay, is generated by neutral top partners in the Folded SUSY~\cite{Burdman:2006tz} and Fraternal Twin Higgs~\cite{Craig:2015pha} models, with masses indicated on the horizontal axes. The solid blue curve shows the reach for 8 decays in the MATHUSLA decay volume, corresponding to the exclusion limit for 50\% reconstruction efficiency expected for near-background-free searches. The dashed curves represent theoretical uncertainties in this reach from unknown aspects of non-perturbative dark $N_f = 0$ QCD.}
\label{f.glueballreach}
\end{figure}

The significance of this sensitivity is vividly demonstrated by considering MATHUSLA's reach for dark glueballs produced in exotic Higgs decays. In practical terms, this can be regarded as a slight elaboration on the minimal exotic Higgs decay simplified model, and is realized in many hidden valleys and, in particular, the 
Fraternal Twin  Higgs~\cite{Craig:2015pha} models and Folded SUSY~\cite{Burdman:2006tz}  solutions to the little Hierarchy problem. The sensitivity of MATHUSLA to these decays for the current 40m geometry is shown in Figure~\ref{f.glueballreach}, as a function of the dark glueball mass and SM-neutral top partner mass. MATHUSLA effectively probes neutral naturalness solutions of the little hierarchy problem in large parts of model's motivated parameter space with neutral top partner masses below a TeV.

\section{Readiness and Expected Challenges}

The MATHUSLA collaboration recently completed a Conceptual Design Report~\cite{Aitken:2025cjq}, and R\&D efforts at the University of Toronto, University of Victoria, Tel Aviv University, and other affiliated partner labs have addressed many basic questions and mostly settled the detailed design of the bar assemblies which are MATHUSLA's basic detector building block. Collaboration with Canadian and CERN engineers also produced engineering concepts for the detector structure and experimental hall near CMS.
A test stand was operated above ATLAS in 2018~\cite{Alidra:2020thg}, which verified basic operational principles and LHC background simulations. 
Two new test stands are currently operating at the Universties of Toronto and Victoria to aid in ongoing R\&D efforts.

One of the R\&D priorities for the immediate future is the detailed design of a MATHUSLA DAQ and trigger system that can supply a L1 signal to the CMS trigger, beyond the 
test stand trials and conceptual studies, respectively, that were conducted for the CDR to demonstrate feasibility. More detailed engineering studies to settle on the precise experimental site near CMS and produce a shovel-ready design for the experimental hall and the detector structure will also be required. 

\textbf{The recent pivot in US funding priorities away from auxiliary LHC experiments makes European support and participation in MATHUSLA, in addition to various already funded efforts and potential future large-scale participation from Canada and Israel, crucial.}

\section{Construction and operational costs}

We briefly summarize the total cost of the experiment as estimated in the CDR~\cite{Aitken:2025cjq}. The construction of the experimental hall, including associated infrastructure like electrical supply, HVAC, etc, is estimated at roughly 25M CHF, taking into account the significantly increased cost associated with building on 'made ground' which comprises the (currently) identified site.
The detector itself is estimated to carry a total cost of roughly 38M CHF, see Table~\ref{tab:cost}, which includes maintenance costs over the time scale of the HL-LHC.

\begin{table}[hpbt]
\centering
\begin{tabular}{lr}
\hline
\multicolumn{1}{r}{}          & \textbf{Total {[}k\$ CAD{]}} \\ \hline
\textbf{Detector Fabrication}  & \textbf{\$19,179}        \\
\hspace{2mm} Scintillator bars               & \$4,914                  \\
\hspace{2mm} WLS fibers                      & \$2,859                  \\
\hspace{2mm} SiPMs                           & \$2,008                  \\
\hspace{2mm} Aluminum casing                   & \$5,556                  \\
\hspace{2mm} Glue                           & \$2,342                  \\
\hspace{2mm} Assembly \& test equipments      & \$1,500                  \\
\textbf{Electronics}           & \textbf{\$10,798}         \\
\hspace{2mm} Frontend                       & \$996                    \\
\hspace{2mm} Data acquisition               & \$5,378                  \\
\hspace{2mm} Cables                          & \$1,793                  \\
\hspace{2mm} Miscellaneous other components                           & \$1,992                  \\
\hspace{2mm} Server \& network             & \$640                    \\
\textbf{Detector installation} & \textbf{\$7,590}     \\
\hspace{2mm} Support structure material cost                & \$2,100     \\
\hspace{2mm} Physical installation labour cost                & \$2,100     \\
\hspace{2mm} Sensor shipping \& handling         & \$3,112     \\
\hspace{2mm} Sensor assembly \& testing    & \$278     \\
\textbf{Detector operations \& project management}    & \textbf{\$6,300}     \\
\hspace{2mm} Salaries                       & \$5,740     \\
\hspace{2mm} Travel                         & \$560     \\ 
\textbf{Maintenance \& repair}    & \textbf{\$1,477}     \\ \hline  
\textbf{Total}                 & \textbf{CAD \$45,344}     \\
\textbf{With 35\% contingency}   & \textbf{CAD \$61,214.}     \\
\hline
\end{tabular}
\label{tab:cost}
\caption{Summary of cost estimation for MATHUSLA-40 in CAD. The total including contingency corresponds to $\approx$ 40M €, or 38M CHF, at the time of writing.}
\end{table}

\section{Timeline}

In the immediate future, MATHUSLA partner labs will engage in the additional R\&D needed to finalize some outstanding aspects of the detector design, including a sufficiently fast trigger system to supply a CMS L1 signal. 
Within 1-2 years of CERN approving the MATHUSLA experiment, international partner labs (likely in Canada, Europe and/or Israel) will begin setting up assembly lines for the construction and quality-testing of \barasss, and produce within one year all the \barasss required for a single \tower. 
These \barasss would then be shipped to CERN and installed in a first \tower over the coming months. 
The first \tower will collect cosmic ray alignment and calibration data for up to a year, during which time partner labs continue to produce  bar assemblies for the other 15 \towers, the \wallveto and the \walltrackingmodules.
Starting in 2029 (at the earliest), the first module will collect LLP search data during HL-LHC runs. Installation of  additional modules will proceed over the course of about a year while existing modules take data as they are completed.

\bibliography{references}
\bibliographystyle{JHEP}

\end{document}